\RequirePackage{ifpdf}
\ifpdf 
\documentclass[pdftex]{sigma}
\else
\documentclass{sigma}
\fi

\numberwithin{equation}{section}

\begin{document}

\allowdisplaybreaks

\renewcommand{\PaperNumber}{084}

\FirstPageHeading

\ShortArticleName{Para-Grassmannian Coherent and Squeezed States for
Pseudo-Hermitian $q$-Oscillator}

\ArticleName{Para-Grassmannian Coherent and Squeezed States\\ for
Pseudo-Hermitian $\boldsymbol{q}$-Oscillator\\ and their Entanglement}

\Author{Yusef MALEKI}

\AuthorNameForHeading{Y.~Maleki}

\Address{Department of Physics, University of Mohaghegh Ardabili, Ardabil, 179, Iran}
\Email{\href{mailto:maleki.thph1365@yahoo.com}{maleki.thph1365@yahoo.com}}

\ArticleDates{Received May 27, 2011, in f\/inal form August 19, 2011;  Published online August 25, 2011}

\Abstract{In this parer, $q$-deformed oscillator for pseudo-Hermitian
systems is investigated and pseudo-Hermitian appropriate coherent
and squeezed states are studied. Also, some basic properties of
these states  is surveyed. The over-completeness property of the
 para-Grassmannian pseudo-Hermitian coherent states
(PGPHCSs) examined, and also the stabi\-li\-ty of  coherent and
squeezed states discussed. The pseudo-Hermitian supercoherent states
as the product of a pseudo-Hermitian bosonic coherent state and a
para-Grassmannian pseudo-Hermitian coherent state  introduced, and
the method also developed to def\/ine pseudo-Hermitian supersqueezed
states. It is also argued that, for $q$-oscillator algebra of $k+1$
degree of nilpotency based on the changed ladder operators, def\/ined
in here, we can obtain deformed $SU_{q^2}(2)$ and $SU_{q^{2k}}(2)$
and also $SU_{q^{2k}}(1,1)$. Moreover, the entanglement of
multi-level para-Grassmannian pseudo-Hermitian coherent state will
be considered. This is done by choosing an appropriate weight
function, and integrating over tensor product of PGPHCSs.}


\Classification{81R30; 15A75; 81P40}

\section{Introduction}
The study of non-Hermitian Hamiltonians with real spectra have
attracted a great deal of attention in the last decade \cite{Bender1,Bender2,Bender3,Mos3,Mos1,Mos2}.
Bender and his coauthors \cite{Bender2,Bender3} found examples of
 non-Hermitian Hamiltonians with real spectra. Since
these Hamiltonians were invariant under ${\mathcal PT}$ transformations,  their
spectral properties were considered to be related to their
${\mathcal PT}$-symmetry.  Later, the basic structure responsible for the reality
of the spectrum of a non-Hermitian Hamiltonian was explored by
Mostafazadeh, and he mentioned that pseudo-Hermiticity is the reason
for the particular spectral properties of these Hamiltonians, he
also investigated its various consequences and its connection with
${\mathcal PT}$-symmetric quantum systems and  pointed out that all the
${\mathcal PT}$-symmetric non-Hermitian Hamiltonians studied in the literature
belong to the class of pseudo-Hermitian Hamiltonians. Consequently,
in the framework of non-Hermitian systems, the notion of
pseudo-Hermitian operator as a non-Hermitian operator with real
spectrum has been found to have a considerable importance~\cite{Mos1,Mos2}.

In the same direction, the big number of works on  coherent states
and their various applications, indicates an excited interest of
 today's researchers in this f\/ield~\cite{perlomov,Fujii1,Naj4,Naj5,Naj3,Wang1,Wang2,Wang3,Wang4,Baz1,Baz2,Baz3,Glauber}.
Among all, Grassmannian coherent states and their
generalizations have been widely investigated in recent years, and
various applications of Grassmann variables have been studied in
theoretical physics and quantum optics~\cite{Baz1,Baz2,Glauber,Mansour,Mansour2}.
In~\cite{Cabra}, the authors focused on the $q$-oscillator algebra
and introduced appropriate coherent states based on para-Grassmann
variables. Recently, some attempts have been made to develop the
study of coherent states to pseudo-Hermitian quantum systems,
pseudo-Hermitian coherent state for bosonic systems, and
pseudo-Hermitian coherent state for two-level and $n$-level systems
based on Grassmann numbers have been studied~\cite{trifb1,trifb2,Naj1}.  In the
present paper, $q$-deformed oscillator for pseudo-Hermitian systems
will be investigated and pseudo-Hermitian appropriate coherent
states will be considered. The pursued method in def\/ining coherent
states  here is inspired by the def\/inition in~\cite{Cabra}. The
over-completeness property of these pseudo-Hermitian
para-Grassmannian pseudo-Hermitian coherent states (PGPHCSs) will be
examined, and using the def\/inition of the coherent states for
pseudo-Hermitian bosonic systems, pseudo-Hermitian supercoherent
states as the product of a pseudo-Hermitian bosonic coherent state
by a pseudo-Hermitian para-Grassmannian coherent state will be
introduced~\cite{Daoud1,Daoud2,Nieto,Baz3}.   The method also will be developed to
pseudo-Hermitian squeezed states and pseudo-Hermitian supersqueezed
states. It will also be argued that for $q$-oscillator algebra of
$k+1$ degree of nilpotency based on the changed ladder operators,
def\/ined in here we can develop deformed $SU_{q^2}(2)$ and
$SU_{q^{2k}}(2)$ and also $SU_{q^{2k}}(1,1)$. One of the aims of the
present work is to consider entanglement of multi-level
PGPHCSs~\cite{Nielsen1,petz1}. This is the generalization of the work
done in~\cite{Naj5,Naj3} to the PGPHCSs. Similar to what we
have done in those papers, choosing some appropriate weight
functions, we conclude that it is possible to construct some
entangled pure states for pseudo-Hermitian systems~\cite{Naj5,Naj3}.
The pseudo-Hermitian version of the  Bell, {\bf{W}} and  {\bf{GHZ}}
will be constructed. Therefore, the PGPHCSs can be applied in
quantum information and quantum computation.

This paper is organized as follows. In Section~\ref{section2}, the concept of a
pseudo-Hermitian operator is introduced  and the basic spectral
properties of pseudo-Hermitian Hamiltonians is given. In Section~\ref{section3},
the def\/inition and basic properties of the para-Grassmann variables
and $q$-oscillator algebra is considered. Then the notion of
pseudo-Hermitian $q$-oscillator algebra and their proper\-ties is given
and also, the appropriate PGPHCSs are introduced. It is shown that
unlike the canonical Hermitian coherent states, PGPHCSs satisfy
bi-over-completeness condition instead of over-completeness one. The
time evolution and  stability of the  PGPHCSs are investigated. Also
the construction of pseudo-Hermitian supercoherent states is
studied. Finally modif\/ied  $SU_{q^2}(2)$ and $SU_{q^{2k}}(2)$ and
also $SU_{q^{2k}}(1,1)$ are introduced.
 In Section~\ref{section4}, the def\/inition of squeezed states and their stability and
 supersqueezed states are argued. Section~\ref{section5} is devoted to the
entanglement of multi-level PGPHCSs. Choosing appropriate weight
functions, we construct some entangled pure states for
pseudo-Hermitian systems. The pseudo-Hermitian version of the Bell,
{\bf{W}} and  {\bf{GHZ}} using two level PGPHCSs, is investigated.
These states degenerate to the well-known  Bell, {\bf{W}} and
{\bf{GHZ}} states, when PGPHCSs reduce to the usual Hermitian
coherent staes. In addition, some examples of entangled qutrit
states are given. Finally, the method is developed for qudit case. A~brief conclusion is given in section~\ref{section6}.

\section{Pseudo-Hermitian operators}\label{section2}

In this section, the notion of the pseudo-Hermiticity of operators
is considered. In fact the necessary and suf\/f\/icient condition for
the reality of the spectrum of a non-Hermitian Hamiltonian admitting
a complete set of biorthonormal eigenvectors is related to their
pseudo-Hermiticity. In order to def\/ine the pseudo-Hermitian
operator, consider $H:\mathcal{{H}}\rightarrow \mathcal{H}$, to be a
linear operator acting in a Hilbert space $\mathcal{{H}}$ and
$\eta:\mathcal{H}\rightarrow \mathcal{H}$, as a linear Hermitian
automorphism. Then the operator
\begin{gather}\label{phh1}
H^\sharp=\eta^{-1}H^\dag\eta,
\end{gather}
is def\/ined as the $\eta$-pseudo-Hermitian adjoint of~$H$. The operator
H is said to be $\eta$-pseudo-Hermitian (with respect to $\eta$) if
$H^\sharp=H$. If H is taken to be a pseudo-Hermitian diagonali\-zab\-le
linear operator then, the eigenvalues are real or come in complex
conjugate pairs and the multiplicity of complex conjugate
eigenvalues are the same (to review the basic properties of
pseudo-Hermitian operators see~\cite{Mos3,Mos1,Mos2}). According to~\cite{Mos1}
considering diagonalizable opera\-tors~$H$ with discrete spectrum, a
complete set of biorthonormal eigenvectors
$\{|\psi_{i}\rangle,|\varphi_i\rangle\}$ exists such that
\begin{gather*}
H|\psi_{i}\rangle={E_i}|\psi_i\rangle, \qquad
H^\dagger|\varphi_i\rangle=\bar{E_i}|\varphi_i\rangle,
\\
\langle\varphi_i|\psi_i\rangle=\delta_{ij},
\qquad
\sum_i|\psi_i\rangle\langle\varphi_i|=\sum_i|\varphi_i\rangle\langle\psi_i|=I.
\end{gather*}
 In fact, any non-Hermitian Hamiltonian
with a discrete real spectrum and a complete biorthonormal system of
eigenbasis vectors is pseudo-Hermitian. Considering the above
discussion, for a~nondegenerate case, the
 operator $\eta$ and it's inverse
satisfy the (\ref{phh1}) by taking
\[
\eta=\sum_i|\varphi_i\rangle\langle\varphi_i|, \qquad
\eta^{-1}=\sum_i|\psi_i\rangle\langle\psi_i|,
\]
thus one has
\[
|\varphi_i\rangle=\eta|\psi_i\rangle, \qquad
|\psi_i\rangle=\eta^{-1}|\varphi_i\rangle.
\]
It is notable that $\eta$ is not unique, however it can be expressed
as above.

\section[Pseudo-Hermitian $q$-oscillator and  coherent states]{Pseudo-Hermitian $\boldsymbol{q}$-oscillator and  coherent states}\label{section3}

\subsection[Para-Grassmann variables and $q$-oscillator algebra]{Para-Grassmann variables and $\boldsymbol{q}$-oscillator algebra}\label{section3.1}

Grassmann variables and their dif\/ferent generalizations
along with
their applications have been widely surveyed and the mathematical
structure of these non-commuting mathematical objects were studied
in resent years \cite{Baz3,Glauber,Cabra,Majid,Kerner,Filip,Isa,Cug,ILInski}. To def\/ine para-Grassmann
variables, consider the non-commutative variables $\theta$ and
$\bar{\theta}$, which satisfy the following relations~\cite{Cabra}
\[
 \theta^{p+1}=0,\qquad \bar{\theta}^{p+1}=0,\qquad
 \theta\bar{\theta}=q^2\bar{\theta}\theta, \qquad \textrm{where} \qquad q^2=e^{\frac{2\pi
 i}{p+1}},
\]
in the above relations, $p$ is a non-zero integer number. The integral
calculus appropriate for these variables is
\[
\int d\theta \theta^{n}=\delta_{n,p}\sqrt{[p]!},\qquad
\textrm{where} \qquad [X]=\frac{q^{2X}-1}{q^2-1}, \qquad
[n]!=[n]\cdots [1].
\]
The $q$-deformed oscillator satisf\/ies the following relations
\cite{Baz3,Baz4,Chaichian}
\[
aa^\dag-qa^\dag a=q^{-N},\qquad  aa^\dag-q^{-1}a^\dag
a=q^{N}.
\]
We note that annihilation and creation operators have the nilpotency
degree of order ${p+1}$.

\subsection[Pseudo-Hermitian $q$-oscillator algebra]{Pseudo-Hermitian $\boldsymbol{q}$-oscillator algebra}\label{section3.2}

In analogy with the above mentioned $q$-oscillator algebra for
Hermitian systems, let us def\/ine pseudo-Hermitian $q$-oscillator
algebra as follows
\[
aa^\sharp-qa^\sharp a=q^{-N},\qquad
aa^\sharp-q^{-1}a^\sharp a=q^{N},
\]
and also, considering the dual space we have
\[ \tilde{a}a^\dag-qa^\dag  \tilde{a}=q^{-N},\qquad
 \tilde{a}a^\dag-q^{-1}a^\dag  \tilde{a}=q^{N},
\]
where $a$ is creation operator for pseudo-Hermitian $q$-deformed
oscillator and can be expressed as follows
\[
a:=\sum_{n=0}^{p} \sqrt{[[n+1]]}
|\psi_{n}\rangle\langle\phi_{n+1}| \quad\Longrightarrow\quad
a|\psi_{n}\rangle=\sqrt{[[n]]}  |\psi_{n-1}\rangle,
\]
and let us def\/ine the creation operator for the pseudo-Hermitian
$q$-deformed oscillator in the Fock space spanned by the basis
$\{|\psi_{0}\rangle\cdots |\psi_{n}\rangle\}$ as
\[
a^\sharp:=\eta^{-1}a^\dag\eta:=\sum_{n=0}^{p} \sqrt{[[n+1]]}
|\psi_{n+1}\rangle\langle\phi_{n}|,
\]
where, we have
\[
[[X]]=\frac{q^X-q^{-X}}{q-q^{-1}}.
\]
Considering the dual space, one can def\/ine the creation and
annihilation operators which act on dual states. In analogy with
annihilation operator $a$, let def\/ine
\[
\tilde{a}:=\eta a\eta^{-1}:=\sum_{n=0}^{p} \sqrt{[[n+1]]}
|\phi_{n}\rangle\langle\psi_{n+1}|.
\]
Hence, $a^\sharp$ is $\eta$-pseudo-adjoint to $a$ and $\tilde{a}$ is
$\eta^{-1}$-pseudo-adjoint to $a^\dag$. So, the operator $a^\dag$
for pseudo-Hermitian $q$-deformed oscillator can be written as
\[
a^\dag:=\sum_{n=0}^{p} \sqrt{[[n+1]]}
|\phi_{n+1}\rangle\langle\psi_{n}|.
\]
Thus one has
\[
a^\sharp a=[[N]]  \quad \Longrightarrow  \quad
[[N]]^\sharp =\eta^{-1}[[N]]^\dag\eta=[[N]],
\]
and by choosing $q^2=e^{\frac{2\pi i}{p+1}}$ we obtain
\[
N=\frac{p+1}{\pi}\arcsin\left(a^\sharp
a\sin\frac{\pi}{p+1}\right).
\]
Hence, the action of the operator $N$ on $|\psi_{n}\rangle$ reads
\[
N|\psi_{n}\rangle=n |\psi_{n}\rangle.
\]
Furthermore, we get the commutation relations between the
annihilation and creation operators and the operator $N$ as
\[
[N,a]=-a, \qquad [N,a^\sharp]=-a^\sharp.
\]
For the dual space operators we can write
\[
a^\dag \tilde{a}=[[N']] \quad \Longrightarrow \quad [[N']]^\sharp
=\eta[[N']]^\dag\eta^{-1}=[[N']].
\]
Again, by choosing $q^2=e^{\frac{2\pi i}{p+1}}$ we obtain
\[
N'=\frac{p+1}{\pi}\arcsin\left(a^\dag
\tilde{a}\sin\frac{\pi}{p+1}\right).
\]

\subsection{Change of variables and coherent states}\label{section3.3}

 Now consider the following change of variables
\[
b=q^{\frac{N}{2}}a, \qquad
\bar{b}=a^{\sharp}q^{\frac{N}{2}} \qquad \textrm{and} \qquad
c=q^{\frac{N'}{2}}\tilde{a}, \qquad
\bar{c}=a^{\dag}q^{\frac{N'}{2}}.
\]
These new operators are necessary for the forthcoming purposes
throughout the paper, also the coherent and squeezed states will be
def\/ined based on these operators not~$a$ and~$a^{\sharp}$ (also~$\tilde{a}$ and~$a^{\dag}$). Taking into account these variables we
see that
\begin{gather}\label{b,bbar}
b \bar{b}-q^2\bar{b}b=1, \qquad \bar{b} b-b \bar{b}=q^{2N} , \qquad
\bar{b} b=[N],
\end{gather}
and for the dual operators we get
\begin{gather}\label{c,cbar}
c\bar{c}-q^2\bar{c}c=1, \qquad \bar{c}c-c \bar{c}=q^{2N},  \qquad
\bar{c} c=[N],
\end{gather}
We not that in~\cite{Naj1} the creation operator  $b^{\sharp}$ is
$\eta$-pseudo-Hermitian adjoint of the annihilation operator~$b$,
but our changed ladder operator $\bar{b}$ is not
$\eta$-pseudo-Hermitian adjoint of $b$, and we have
$b^{\sharp}=\bar{b}q^{-N}$. One could easily check that the
pseudo-Hermitian number states can be expressed in terms of~$b$ and~$\bar{b}$ (also~$c$ and~$\bar{c}$ for the dual space), and these
operators annihilate and create the number states as follows
\[
|\psi_{n}\rangle=\frac{( \bar{b})^n}{
\sqrt{[n]!}}|\psi_{0}\rangle, \qquad b|\psi_{n}\rangle=\sqrt{[n]}
|\psi_{n-1}\rangle, \qquad \bar{b}|\psi_{n}\rangle=\sqrt{[n+1]}
|\psi_{n}\rangle.
\]
Thus
\[
[N,b]=-b, \qquad [N,\bar{b}]=\bar{b},  \qquad [N,c]=-c,
\qquad [N,\bar{c}]=\bar{c}.
\]
For the future uses, we need to consider the commutation relation
between these operators and para-Grassmannian variables
\[
[\theta,b]_{q^{2}}=0,\qquad
\qquad[\bar{b},\theta]_{q^{2}}=0,\qquad
[\bar{\theta},\bar{b}]_{q^{2}}=0,\qquad
[b,\bar{\theta}]_{q^{2}}=0,
\]
where the $p$-commutator is def\/ined as
\[
[A,B]_{p}=AB-p BA.
\]
Let us have the following quantization relations between the
biorthonormal pseudo-Hermitian eigenstates $\{|\psi_{n}\rangle,
|\phi_{n}\rangle\}$ and para-Grassmannian variables
 \begin{alignat*}{3}
&  \theta   |\psi_{n}\rangle = q^{^{-2n}}   |\psi_{n}\rangle
 \theta, \qquad &&  \bar{\theta}  \langle\psi_{n}|  = q^{^{-2n}}   \langle\psi_{n}|  \bar{\theta}, & \\
&  \theta   \langle\psi_{n}|  = {q}^{^{2n}}
\langle\psi_{n}|  \theta,\qquad &&  \bar{\theta}   |\psi_{n}\rangle
= {q}^{^{2n}}   |\psi_{n}\rangle
\bar{\theta}, & \\
&  \theta   |\phi_{n}\rangle   =q^{^{-2n}}   |\phi_{n}\rangle
 \theta,\qquad &&  \bar{\theta}   \langle\phi_{n}|   = q^{^{-2n}}    \langle\phi_{n}|  \bar{\theta}, &
 \\
&  \theta   \langle\phi_{n}|  = {q}^{^{2n}}
\langle\phi_{n}|   \theta , \qquad && \bar{\theta}   |\phi_{n}\rangle
  = {q}^{^{2n}}   |\phi_{n}\rangle   \bar{\theta}.&
\end{alignat*}
 Now let $|\theta\rangle$ be a PGPHCS, then following the def\/inition of the canonical Hermitian coherent
 state,
  the  PGPHCS  $|\theta\rangle$ can be def\/ined as the eigenstates of annihilation operator
  $b$ for the space spanned by the basis  $\{|\psi_{n}\rangle\}$.
  Therefore, PGPHCS  $|\theta\rangle$ reads
\begin{gather}\label{coh1}
b  |\theta\rangle=\theta   |\theta\rangle.
\end{gather}
In order to f\/ind the explicit form of the   PGPHCS $|\theta\rangle$,
we can generically write
\begin{gather}\label{coh2}
|\theta\rangle=\sum^p_{n=0}c_n \theta^{n}|\psi_{n}\rangle.
\end{gather}
The two equations (\ref{coh1}) and (\ref{coh2}) give
\begin{gather}\label{coh-gen}
|\theta\rangle=\sum^p_{n=0}\frac{q^{n(n+1)}}{\sqrt{[n]}!}
\theta^{n}|\psi_{n}\rangle=e^{\bar{b}\theta}_q|\psi_{0}\rangle,
\end{gather}
we can interpret $D(\theta)=e_q^{\bar{b}\theta}$ as the displacement
operator for the PGPHCS. We note that \textit{$q$-deformed}
exponential is
\[
e_q^{x}=\sum_{n=0}^p\frac{x^n}{[n]!}.
\]
We can easily check that the action of the creation operator
$\bar{b}$ on the coherent state gives
\[
\bar{b}|\theta\rangle=q^{-2}\sum^p_{n=1}q^{n(n+1)}\frac{[n]}{\sqrt{[n]!}}
\theta^{n-1}|\psi_{n}\rangle.
\]
The overlap between the state $|\phi_{n}\rangle$ and the coherent
state is given as
\begin{gather}\label{overlap}
\langle\phi_{n}|\theta\rangle=\frac{q^{-n(n-1)}}{\sqrt{[n]}!}
\theta^{n}.
\end{gather}
We can also def\/ine the other set of coherent states based on the
creation operator $\bar{b}$ with para-Grassmanian eigenvalue as
follows
\[
\langle\bar{\theta}|\bar{b}=\langle\bar{\theta}|\bar{\theta}.
\]
Thus, in the terms of number state basis the  coherent states of the
creation operator $\bar{b}$ can be expanded as
\[
\langle\bar{\theta}|=\sum^p_{n=0}\frac{q^{n(n-1)}}{\sqrt{[n]}!}
\bar{\theta}^{n}\langle\psi_{n}|.
\]
Similar to (\ref{overlap}), the overlap between the state
$|\phi_{n}\rangle$ and this coherent state depends on $n$ and takes
the following form
\[
\langle\bar{\theta}|\phi_{n}\rangle=\frac{q^{n(n-1)}}{\sqrt{[n]}!}
\bar{\theta}^{n}.
\]
As was seen, for the pseudo-Hermitian systems, there are two sets of
annihilation and creation operators, with respect to the basis
$\{|\psi_{n}\rangle\}$ and $\{|\phi_{n}\rangle\}$ respectively.
Therefore, a coherent state can be def\/ined in the space spanned by
the basis $\{|\phi_{n}\rangle\}$, which is the eigenstates of
annihilation operator $c$ with para-Grassmanian eigenvalue $\theta$
similar to what we had for the operator $b$, as follows
\[
c  \widetilde{|\theta\rangle}=\theta\ \widetilde{ |\theta\rangle}.
\]
Again taking the general form of the coherent state in (\ref{coh2})
and using the above relation for the coherent states, it is
straightforward to see that the coherent state is
\[
\widetilde{|\theta\rangle}=\sum^p_{n=0}\frac{q^{n(n+1)}}{\sqrt{[n]}!}
\theta^{n}|\phi_{n}\rangle=e^{\bar{c}\theta}_q|\phi_{0}\rangle.
\]
Moreover, by acting the creation operator $\bar{c}$ on coherent
state we get
\[
\bar{c}|\theta\rangle=q^{-2}\sum^p_{n=1}q^{n(n+1)}\frac{[n]}{\sqrt{[n]!}}
\theta^{n-1}|\phi_{n}\rangle.
\]
The overlap between the state $|\psi_{n}\rangle$ and the coherent
state of the annihilation operator $c$ is proportional to
$\theta^{n}$ as
\[
\langle\psi_{n}\widetilde{|\theta\rangle}=\frac{q^{-n(n-1)}}{\sqrt{[n]}!}
\theta^{n}.
\]
Hence, the overlap between the state $|\phi_{n}\rangle$ and
(\ref{coh-gen}) is equal to the overlap between the state~$|\psi_{n}\rangle$ and the coherent state of the annihilation
operator $c$ def\/ined above, i.e.,
$\langle\bar{\theta}|\phi_{n}\rangle=\langle\psi_{n}\widetilde{|\theta\rangle}$.
Furthermore, one may def\/ine the coherent state of the creation
operator $\bar{c}$, as the superposition of the number states of the
dual space with para-Grassmanian eigenvalue $\bar{\theta}$ as
\[
\widetilde{\langle\bar{\theta}|}\bar{c}=\widetilde{\langle\bar{\theta}|}\bar{\theta}.
\]
Again we can expand the coherent state based on the number states of
the related space and powers of the para-Grassmanian
variable~$\bar{\theta}$ as
\[
\widetilde{\langle\bar{\theta}|}=\sum^p_{n=0}\frac{q^{n(n-1)}}{\sqrt{[n]}!}
\bar{\theta}^{n}\langle\phi_{n}|.
\]
Hence, we have
\[
\widetilde{\langle\bar{\theta}|}\psi_{n}\rangle=\frac{q^{n(n-1)}}{\sqrt{[n]}!}
\bar{\theta}^{n}.
\]
Consequently, there is a relation between the overlaps of coherent
states with number states as
$\langle\bar{\theta}|\phi_{n}\rangle=\widetilde{\langle\bar{\theta}|}\psi_{n}\rangle$.
So, there is a relation between the overlaps of the coherent states
of the annihilation operators with number states of their dual
states and also between the overlaps of the coherent states of the
creation operators with number states of their dual states as well.

\subsection{Resolution of identity}\label{section3.4}

Here, the over-completeness property of the states  $|\theta\rangle$
and $\widetilde{|{\theta}\rangle}$  is examined. By introducing the
generic form of the weight function as
\[
w(\theta, \bar{\theta})=\sum_{k,l=0}^{n-1} c_{kl}\theta^k
\bar{\theta}^l.
\]
We see that the integrals of the terms
$|\theta\rangle\langle\theta|$ and
$\widetilde{|{\theta}\rangle}\widetilde{\langle{\theta}|}$ do not
satisfy over-completeness relation. Hence, in order to realize the
resolution of identity, it is necessary to consider  the
biorthonormal nature of the pseudo-Hermitian systems. So, as done in~\cite{trifb2,Naj1}, it is reasonable to check the integrals
$|{\theta}\rangle\widetilde{\langle{\theta}|}$ and
$\widetilde{|{\theta}\rangle}\langle{\theta}|$  with the measure
$d\bar{\theta}  d\theta   w(\theta, \bar{\theta})$. Choosing the
proper weight function, we can  write the resolution of identity as
\begin{gather}\label{resolution2}
\int d\bar{\theta}  d\theta   w(\theta, \bar{\theta})
|\theta\rangle\widetilde{\langle{\theta}|} = \int d\bar{\theta}
d\theta  w(\theta, \bar{\theta})
\widetilde{|{\theta}\rangle}\langle\theta|=I.
\end{gather}
To obtain the explicit form of the weight function we put
$|\theta\rangle$ and $\widetilde{\langle\theta|}$ into
(\ref{resolution2}) and get
\[
\int d\bar{\theta}  d\theta   w(\theta, \bar{\theta})\
|\theta\rangle\widetilde{\langle{\theta}|} =\int d\bar{\theta}
d\theta  \sum_{k,l=0}^{p}c_{k,l}\theta^k
\bar{\theta}^l\sum_{n,m=0}^{p}\frac{{q}^{{n(n+1)}}}{\sqrt{[{n}]!}}
\theta^n|\psi_{n}\rangle\langle\phi_{m}|  \bar{\theta}^m
\frac{{\bar{q}}^{{m(m+1)}}}{[{m}]!}.
\]
Accounting for the quantization
 and the integration rules of
para-Grassmann variables, and utilizing the completeness of the
biorthonormal basis of the pseudo-Hermitian  system, i.e.,
$\sum\limits_{n=0}^{p}|\psi_{n}\rangle\langle\phi_{n}|=I $, we can
determine the coef\/f\/icients $c_{kl}$  as
\[
c_{k,l}=\frac{[p-l]!}{[p]!}  q^{-2l(l+1)} \delta_{k,l}.
\]
Thus the weight function reads
\[
w(\theta, \bar{\theta})=\sum_{l=0}^{p} \frac{[p-l]!}{[p]!}
q^{-2l(l+1)}   \theta^l \bar{\theta}^l.
\]
Consequently the coherent states $|\theta\rangle$ and
$\widetilde{|{\theta}\rangle}$ provide the bi-over-completeness
relation for pseudo-Hermitian systems of biorthnormal coherent
states.  Note that when the pseudo-Hermitian system reduces to the
Hermitian one, the coherent state $\widetilde{|{\theta}\rangle}$
degenerates to $|\theta\rangle$. It is in accordance with the
over-completeness property  of the Hermitian systems.

\subsection{Time evolution of PGPHCS}\label{section3.5}

 In this subsection, the condition for the stability of the PGPHCS will be
investigated. It will be shown that the coherent states remain
eigenstate of the annihilation operator~$b$, for any time evolution
of the initial coherent state $|\theta,0\rangle \equiv
|\theta\rangle$, i.e., $|\theta,t\rangle$, taking a proper
condition. This means that
\[
b  |\theta,t\rangle=\theta(t)  |\theta,t\rangle,\qquad \textrm{where}
\qquad|\theta,t\rangle=e^{-iHt}|\theta\rangle.
\]
Taking the explicit form of the coherent state  $|\theta\rangle$,
 one can write
\[
|\theta,t\rangle=\sum_{k=0}^{p}
\frac{{q}^{^{{k(k+1)}}}}{\sqrt{[{k}]!}} \theta ^{k}
e^{-iE_{k}t}|\psi_{k}\rangle.
\]
Now we note that, since $k$ changes, all the energies related to the
various  $k$s form the set $\{E_{0},E_{1},\dots,E_{p}\}$, which are
real parameters. Thus, in general we can write
\[
E_{k}=(kc_k+1)E_{0}.
\]
Now, if $c_k$ is taken to be constant ($c_i=c_j=c$ ), the evolved
coherent state reads
\[
|\theta,t\rangle=e^{-iE_{0}t} |\theta(t)\rangle, \qquad
\textrm{where} \qquad \theta(t)=e^{-icE_0t}\theta.
\]
So, we obtain the condition that  the evolved coherent state remain
eigenstate of the annihilation operator~$b$. This result shows that
the  time evolved pseudo-Hermitian $q$-deformed oscillator coherent
state $|\theta\rangle$ is stable. The similar discussion holds for
$|\widetilde{\theta,t}\rangle$'s, and we can also see that
$|\theta,t\rangle$ and $|\widetilde{\theta,t}\rangle$ satisfy the
resolution of identity.

\subsection{Supercoherent states}\label{section3.6}

Supersymmetric quantum mechanics needs two degrees of freedom: one
described by a complex variable and the other described by a
Grassmann variable. Here, the notion of pseudo-Hermitian
supercoherent states will be developed. In order to provide a
suitable supersymmetric Hamiltonian, we can follow the results of
supersymmetric Hamiltonian for Hermitian systems based on
 generalized Weyl--Heisenberg algebra~\cite{Daoud1,Daoud2,Daoud3}. In what follows, the pseudo-Hermitian
 supersymmetric Hamiltonian (PHSUSH) for two level systems will be
 established~\cite{Cherbal3}. To do so, we use  pseudo-boson
annihilation and creation operators (see~\cite{trifb1}). According
to~\cite{trifb1}, if~$\texttt{a}$ and~$\texttt{a}^{\dag}$ are taken
to be usual annihilation and creation operators for a bosonic
system, then the $\eta'$-pseudo-Hermitian adjoint of  $\texttt{a}$
is $\texttt{a}^\sharp:=\eta'^{-1}\texttt{a}^{\dag}\eta'$, and also
$\tilde{\texttt{a}}:=\eta' \texttt{a}\eta'^{-1}$. On the other hand,
for a two level system we have $b \bar{b}+\bar{b}b=1$ ($b^2=\bar{b}^2=0$ and $q^2=-1$). This is similar to pseudo-fermionic
algebra~\cite{trifb2}. In other words, by replacing $\bar{b}$ with~$b^\sharp$, the pseudo-fermionic algebra can be obtained ($b
b^\sharp+b^\sharp b=1$). We note that the operators
are changed, but the obtained coherent states for both systems can
be the same. Therefore, based on pseudo-fermionic and
pseudo-bosonic operators, one may def\/ine the PHSUSH as
\[
H_1=\texttt{a}^\sharp\texttt{a}+b^\sharp b.
\]
Regarding the fact that the bosonic operators commute with two level
operators and vice versa, this Hamiltonian is
$(\eta'\eta)$-pseudo-Hermitian. Now, taking $Q=\texttt{a}^\sharp b$
and $Q^\sharp=\texttt{a} b^\sharp$, then we have
\[
H_1=QQ^\sharp+Q^\sharp Q.
\]
both of the operators $Q$ and $Q^\sharp$  commute with Hamiltonian.
To generalize the result to multi-level systems, we can replace
$\bar{b}$ with $b^\sharp$ in general, and get the pseudo-Hermitian
version of the deformed algebra $b b^{\dag}-q^2b^{\dag}b=1$
discussed in~\cite{Daoud1,Daoud2,Daoud3}. Again,  the operators are changed, but the
obtained coherent states for both systems can be the same. To see
this we take
$b:=\sum\limits_{n=0}^{p} \sqrt{[n+1]}
|\psi_{n}\rangle\langle\phi_{n+1}|$.
 Therefore, taking
$\bar{b}\equiv b^\sharp$ the supercohertent states can be developed.

The coherent
states for the pseudo-Hermitian boson systems are def\/ined as the eigenstates of the corresponding pseudo-boson
annihilation operators
\[
\texttt{a}|\alpha\rangle=\alpha|\alpha\rangle,
\]
where $\alpha$ is a complex number, and $a$ is the annihilation
operator for pseudo-Hermitian bosonic harmonic oscillator. Thus,
considering the def\/inition of the coherent state in the number
states space, it can be written as follows
\[
|\alpha\rangle=e^{\frac{-|\alpha|^2}{2}}
\sum_{n=0}^\infty\frac{\alpha^n}{\sqrt{n!}}|\psi_{n}\rangle=D(\alpha)|\psi_{0}\rangle,
\]
here, $D(\alpha)$ is the displacement operator
\[
D(\alpha):=\exp(\alpha \texttt{a}^\sharp-\alpha^*\texttt{a}).
\]
On the other hand for the dual space, we can def\/ine the associated
pseudo-bosonic  coherent state using the corresponding annihilation
operator $\tilde{a}$ as
\[
\tilde{\texttt{a}}|\alpha\rangle'=\alpha|\alpha\rangle'.
\]
Again, in the number states space, the coherent state can be
expanded as follows
\[
|\alpha\rangle'=e^{\frac{-|\alpha|^2}{2}}
\sum_{n=0}^\infty\frac{\alpha^n}{\sqrt{n!}}|\phi_{n}\rangle=D'(\alpha)|\phi_{0}\rangle,
\]
where $D'(\alpha)$ is the displacement operator and is similar to
the displacement operator def\/ined for its dual space
\[
D'(\alpha):=\exp(\alpha \texttt{a}^\dag-\alpha^*\tilde{\texttt{a}}).
\]
In what follows the supercoherent state for pseudo-Hermitian system
is def\/ined, using the above pseudo-bosonic  coherent states and
para-Grassmannian coherent states
 for the pseudo-Hermitian $q$-oscillator as follows
\[
|\alpha,\theta\rangle=|\alpha\rangle\otimes|\theta\rangle=D(\alpha,\theta)|\psi_{0}\rangle\otimes|\psi_{0}\rangle,
\]
where, the displacement operator $D(\alpha,\theta)$, is
 the tensor product of displacement operators for bosonic and para-Grassmanian coherent states
\[
D(\alpha,\theta)=D(\alpha)\otimes D(\theta)\quad \Longrightarrow\quad
D(\theta)= e^{\bar{b}\theta}_q.
\]
Thus  we can rewrite the above super coherent state in terms of the
number states as
\[
|\alpha,\theta\rangle=e^{\frac{-|\alpha|^2}{2}}
\sum_{n=0}^\infty\frac{\alpha^n}{\sqrt{n!}}|\psi_{n}\rangle\otimes\sum^p_{n=0}\frac{q^{n(n+1)}}{\sqrt{[n]}!}
\theta^{n}|\psi_{n}\rangle.
\]
The state $|\alpha,\theta\rangle$ is an eigenstate of the operator
$a {b}$ with the eigenvalue $\alpha\theta$. Since for the
pseudo-Hermitian systems we deal with two sets of basis, unlike the
Hermitian coherent states, we have some other possibilities in order
to def\/ine supercoherent states for these systems. For instance, we
can def\/ine the following supercoherent state
\[
|\alpha,\widetilde{\theta}\rangle=|\alpha\rangle\otimes\widetilde{|\theta\rangle}=D(\alpha,\widetilde{\theta})|\psi_{0}\rangle\otimes|\phi_{0}\rangle,
\]
where  the displacement operator $D(\alpha,\widetilde{\theta})$
reads
\[
D(\alpha,\widetilde{\theta})=D(\alpha)\otimes
\widetilde{D(\theta)}\quad \Longrightarrow \quad \widetilde{D(\theta)}=
e^{\bar{c}\theta}_q.
\]
In this case, we have used the para-Grassmanian coherent state of
the dual basis, so in terms of the number states we have
\[
|\alpha,\widetilde{\theta}\rangle=e^{\frac{-|\alpha|^2}{2}}
\sum_{n=0}^\infty\frac{\alpha^n}{\sqrt{n!}}|\psi_{n}\rangle\otimes\sum^p_{n=0}\frac{q^{n(n+1)}}{\sqrt{[n]}!}
\theta^{n}|\phi_{n}\rangle.
\]
Likewise, one may take
\[
|\alpha',\theta\rangle=|\alpha\rangle'\otimes|\theta\rangle=D(\alpha',\theta)|\phi_{0}\rangle\otimes|\psi_{0}\rangle,
\]
similar to the previous cases the displacement operator is
\[
D(\alpha',\theta)=D'(\alpha)\otimes D(\theta),
\]
consequently, we have
\[
|\alpha',\theta\rangle=e^{\frac{-|\alpha|^2}{2}}
\sum_{n=0}^\infty\frac{\alpha^n}{\sqrt{n!}}|\phi_{n}\rangle\otimes\sum^p_{n=0}\frac{q^{n(n+1)}}{\sqrt{[n]}!}
\theta^{n}|\psi_{n}\rangle.
\]
Finally, taking the coherent states def\/ined on the basis
$\{|\phi_{n}\rangle\}$ for both bosonic and para-Grassmanian
coherent state, we obtain the following supercoherent state
\[
|\alpha',\widetilde{\theta}\rangle=|\alpha\rangle'\otimes\widetilde{|\theta\rangle}=D(\alpha',\widetilde{\theta})|\phi_{0}\rangle\otimes|\phi_{0}\rangle,
\]
similarly, by def\/ining
$D(\alpha',\widetilde{\theta})=D'(\alpha)\otimes|\widetilde{\theta}\rangle$,
we can write the coherent state in terms of the number states as
\[
|\alpha',\widetilde{\theta}\rangle=e^{\frac{-|\alpha|^2}{2}}
\sum_{n=0}^\infty\frac{\alpha^n}{\sqrt{n!}}|\phi_{n}\rangle\otimes\sum^p_{n=0}\frac{q^{n(n+1)}}{\sqrt{[n]}!}
\theta^{n}|\phi_{n}\rangle.
\]

\subsection[Deformed  $SU(2)_p$ and $SU(1,1)_p$  for multi-level systems]{Deformed  $\boldsymbol{SU(2)_p}$ and $\boldsymbol{SU(1,1)_p}$  for multi-level systems}\label{section3.7}

In this subsection, using the operators $b$ and $\bar{b}$, the
condition to have deformed~$SU(2)$, i.e.~$SU_p(2)$, will be
considered. It will be shown that, regarding special value for~$p$
which will be derived here, these operators can be generators of
$su_p(2)$ Lie algebra. Note that there is no condition on $p$ and it
is a complex number. Now let us def\/ine
\[
b_z:=[b,\bar{b}]_p=b \bar{b}-p\bar{b}b.
\]
Considering the commutation relation between $b$ and $b_z$
\[
[b_z,{b}]_p=\big[\big(1+p^2-pq^2-p\big)+\big(q^2+p^2q^2-\big(q^4+1\big)p\big)\bar{b}b)\big]b,
\]
to have appropriate solution, it must be proportional to the
operator $b$, so we get
\[
q^2p^2-\big(q^4+1\big)p+q^2=0\quad  \Longrightarrow \quad p=q^2 \qquad
\textrm{or} \qquad p=\bar{q}^{^2}.
\]
Consequently one has
\[
 p=q^2\quad\Longrightarrow \quad[b_z,{b}]_p=[b_z,{b}]_{q^2}=\big(1-q^2\big)b,
\]
and
\[
 p=\bar{q}^{^2}\quad\Longrightarrow \quad[b_z,{b}]_p=[b_z,{b}]_{\bar{q}^{^2}}=(\bar{q}^{^4}-\bar{q}^{^2})b.
\]
From the commutator of  $\bar{b}$ and  ${b}_z$ the same condition
will be required to get proper solution, therefore we get the
following $su_p(2)$ deformed algebras
\[
 [b,\bar{b}]_{q^2}=b_{z}, \qquad
  {[b_{z},b]}_{q^2}=\big(1-q^2\big)b,\qquad
  {[\bar{b},b_{z}]}_{q^2}=\big(1-q^2\big)\bar{b},
\]
and also
\[
[b,\bar{b}]_{\bar{q}^{^2}}=b_{z}, \qquad
 {[b_{z},b]}_{\bar{q}^{^2}}=\big(\bar{q}^{^4}-\bar{q}^{^2}\big)b,\qquad
  {[\bar{b},b_{z}]}_{\bar{q}^{^2}}=\big(\bar{q}^{^4}-\bar{q}^{^2}\big)\bar{b}.
\]
Obviously, if $b^{k+1}=\bar{b}^{k+1}=0$, then $(q^2)^{k+1}=1$, so
there are two dif\/ferent algebras associated with~$p=q^2$ and~$p=q^{2k}$. Therefore, the second algebra can be written as
\begin{gather}
 [b,\bar{b}]_{q^{2k}}=b_{z}, \qquad
  {[b_{z},b]}_{q^{2k}}=\big(q^{2(k-1)}-q^{2k}\big)b,\qquad
 {[\bar{b},b_{z}]}_{q^{2k}}=\big(q^{2(k-1)}-q^{2k}\big)\bar{b}.\label{su3}
\end{gather}
Now let def\/ine the operator $b_z$ in a dif\/ferent way as
\[
b_z:=[\bar{b},b]_p=\bar{b}b -pb\bar{b}.
\]
The commutation relation between $b$ and $b_z$ becomes
\[
[b_z,{b}]_p=\big[\big(p^2+q^2p^2-2p\big)+\big(p^2q^4-2pq^2+1\big)\bar{b}b\big]b,
\]
to have proper solution, the commutation relation must be
independent from the operator $\bar{b}b$, therefore, the coef\/f\/icient
of the operator $\bar{b}b$ must vanish. this means
\[
p^2q^4-2pq^2+1=0\quad  \Longrightarrow \quad \big(pq^2-1\big)^2=0 \quad
\Longrightarrow \quad p=\bar{q}^{^2}.
\]
Thus, in this case we have unique solution for $p$, and the above
commutation relation reduces to
\[
 p=\bar{q}^{^2}\quad\Longrightarrow \quad[b_z,{b}]_p=[b_z,{b}]_{\bar{q}^{^2}}=\big(\bar{q}^{^4}-\bar{q}^{^2}\big)b.
\]
Just like the previous one, from the commutator of  $\bar{b}$ and
${b}_z$ the same condition will be required in order to get proper
solution. If $(q^2)^{k+1}=1$, then we have $p=q^{2k}$. Consequently
the algebra degenerates to
\[
 [\bar{b},b]_{q^{2k}}=b_{z},\qquad
  {[b_{z},b]}_{q^{2k}}=\big(q^{2(k-1)}-q^{2k}\big)b,\qquad
  {[\bar{b},b_{z}]}_{q^{2k}}=\big(q^{2(k-1)}-q^{2k}\big)\bar{b}.
\]
It must be stressed that, although this algebra is similar to
(\ref{su3}), they  correspond to dif\/ferent def\/initions of the
operator $b_z$ and are not the same algebras. The dif\/ference between
def\/inition of the operator $b_z$ in the two algebras may recall the
dif\/ference between the def\/inition of the~$su(2)$ and~$su(1,1)$
algebras. Therefore, one may call this algebra deformed
$su(1,1)_{q^{2k}}$ algebra. It is notable that, the above algebras
were derived based on the relation between the operators~$b$ and~$\bar{b}$, and it would not be possible to develop similar algebras
using the creation and annihilation operators~$a$ and~${a}^\sharp$.

\section{Squeezed states}\label{section4}

By recalling the def\/inition of the standard bosonic harmonic
oscillator squeezing operator, it is worthwhile to def\/ine
para-Grassmannian pseudo Hermitian squeezing operator based on the
new operators for $q$-deformed oscillator as follows
\[
S(\theta)=\exp\left[\frac{1}{2}\big(\theta \bar{b}^{2}-  \bar{\theta}b^2\big)\right].
\]
Therefore, by applying the operator $S(\theta)$ on the vacuum state
$|\psi_{0}\rangle$, the para-Grassmannian pseudo Hermitian squeezed
states can be def\/ined as
\[
|\xi\rangle=S(\theta)|\psi_{0}\rangle.
\]
It is notable that the above operator is squeezing operator for any
$n>2$ level systems, so for three level system (since the operators
$b^{3}$ and $\bar{b}^{3}$ become zero), the squeezing operator can
be expanded as
\[
S(\theta)=I+\frac{1}{2}\big(\theta \bar{b}^{2}-  \bar{\theta}
b^2\big)-\frac{{\theta\bar{\theta}}}{8}
\big(q^{2}\bar{b}^{2}b^2+b^2\bar{b}^{2}\big).
\]
Thus, the three level squeezed state reads
\[
|\xi\rangle=|\psi_{0}\rangle+\frac{\sqrt{[{2}}]}{2}\theta|\psi_{2}\rangle
-\frac{[{2}]}{8}\theta\bar{\theta}|\psi_{0}\rangle
=\left(1-\frac{[{2}]}{4}\theta\bar{\theta}\right)|\psi_{0}\rangle+
\frac{\sqrt{[{2}]}}{2}\theta|\psi_{2}\rangle.
\]
Besides, using the dual operators $c$ and $\bar{c}$, one may
consider the other set of the squeezing operators for
para-Grassmannian pseudo Hermitian $q$-deformed systems as below
\[
\widetilde{S}(\theta)=\exp\left[\frac{1}{2}\big(\theta \bar{c}^{2}-
\bar{\theta}c^2\big)\right],
\]
for three level system, the operators $c^{3}$ and $\bar{c}^{3}$ are
zero, so the squeezing operator $\widetilde{S}(\theta)$ becomes
\begin{gather}\label{squoper2}
\widetilde{S}(\theta)=I+\frac{1}{2}\big(\theta \bar{c}^{2}-
\bar{\theta} c^2\big)-\frac{{\theta\bar{\theta}}}{8}
\big(q^{2}\bar{c}^{2}c^2+c^2\bar{c}^{2}\big).
\end{gather}
Hence, the squeezed state is given by
\[
|\tilde{\xi}\rangle=\widetilde{S}(\theta)|\phi_{0}\rangle.
\]
Using (\ref{squoper2}) we have
\[
|\tilde{\xi}\rangle=|\phi_{0}\rangle+\frac{\sqrt{[{2}}]}{2}\theta|\phi_{2}\rangle
-\frac{[{2}]}{8}\theta\bar{\theta}|\phi_{0}\rangle
=\left(1-\frac{[{2}]}{4}\theta\bar{\theta}\right)|\phi_{0}\rangle+
\frac{\sqrt{[{2}]}}{2}\theta|\phi_{2}\rangle.
\]
It is notable that, the dual squeezing operator
$|\tilde{\xi}\rangle$ can be obtained from the action of $\eta$ on
$|\xi\rangle$. This means
\[
|\tilde{\xi}\rangle=\eta|\xi\rangle=\left(1-\frac{[{2}]}{4}\theta\bar{\theta}\right)|\phi_{0}\rangle+
\frac{\sqrt{[{2}]}}{2}\theta|\phi_{2}\rangle.
\]

\subsection{Time evolution  of squeezed states}\label{section4.1}

In analogy with coherent states, it would be worthy to study the
time evolution of squeezed states. The time evolution of a system is
governed by its Hamiltonian. Here, the time evolution of the three
level squeezed states will be considered and it will be shown that
the three level squeezed state remains squeezed state temporally.
For a three level system we have obtained the squeezed state as
\[
|\xi\rangle=\left(1-\frac{[{2}]}{4}\theta\bar{\theta}\right)|\psi_{0}\rangle+
\frac{\sqrt{[{2}]}}{2}\theta|\psi_{2}\rangle.
\]
Now, applying the time evolution operator on this state we have
\[
|\xi,t\rangle=e^{-iHt}|\xi\rangle=\left(1-\frac{[{2}]}{4}\theta\bar{\theta}\right)e^{-iE_{0}t}|\psi_{0}\rangle+
\frac{\sqrt{[{2}]}}{2}\theta e^{-iE_{2}t}|\psi_{2}\rangle,
\]
or equivalently, it can be rewritten as
\[
|\xi,t\rangle=e^{-iE_{0}t} {S(\theta(t))}|\psi_{0}\rangle, \qquad
\textrm{where} \qquad \theta(t)= e^{-i(E_{2}-E_{0})t}\theta.
\]
Consequently, the time evolved state $|\xi,t\rangle$ is squeezed
state. We note that, without loss of generality, one may take
$E_{0}=0$, and write the time evolved squeezed state $|\xi,t\rangle$
as the action of the operator ${S(\theta(t))}$ on vacuum state
$|\psi_{0}\rangle$. Similarly, for the dual space it can be deduced
that, the time evolved squeezed state is
\[
|\tilde{\xi}\rangle=e^{-iHt}|\tilde{\xi}\rangle=e^{-iE_{0}t}
\widetilde{S}(\theta)|\phi_{0}\rangle, \qquad \textrm{where} \qquad
\theta(t)= e^{-i(E_{2}-E_{0})t}\theta.
\]

\subsection{Supersqueezed states}\label{section4.2}
Using the creation and annihilation operators for pseudo-Hermitian
bosonic system we can def\/ine pseudo-Hermitian squeezing operator in
analogy with the Hermitian  squeezing operator. Hence, let us def\/ine
the pseudo-bosonic squeezing operator as para-Grassmannian
\[
S(\eta)=\exp\left[\frac{1}{2}\big(\eta {a^\sharp}^{2}- {\eta^*}a^2\big)\right],
\]
where $\eta$ is a complex number, and $a$, $a^\sharp$ are bosonic
pseudo-Hermitian operators. Hence the squeezed state for bosonic
pseudo-Hermitian system becomes
\[
|\eta\rangle=S(\eta)|\psi_{0}\rangle=\exp\left[\frac{1}{2}\big(\eta
{a^\sharp}^{2}-\ {\eta^*}a^2\big)\right]|\psi_{0}\rangle.
\]
In addition
\[
|\eta\rangle'=S'(\eta)|\phi_{0}\rangle=\exp\left[\frac{1}{2}\big(\eta
{a^\dag}^{2}-\ {\eta^*}\tilde{a}^2\big)\right]|\phi_{0}\rangle.
\]
Considering these states, and the squeezed states introduced for
para-Grassmann variables one may study the structure of the supersymmetric squeezed states as the tensor product of the
pseudo-bosonic and para-Grassmannian squeezed state
\[
|\eta,\theta\rangle=|\eta\rangle\otimes|\zeta\rangle=S(\eta,\zeta)|\psi_{0}\rangle\otimes|\psi_{0}\rangle,
\]
where the supersqueezing operator is
\[
S(\eta,\theta)= S(\eta)\otimes S(\theta)=\exp\left[\frac{1}{2}\big(\eta
{a^\sharp}^{2}- {\eta^*}a^2\big)\right]\otimes\exp\left[\frac{1}{2}\big(\theta
\bar{b}^{2}- \bar{\theta}b^2\big)\right].
\]
As  mentioned for supercoherent states, there are some other
possibilities in order to def\/ine supersqueezed states for the
pseudo-Hermitian systems. Since, one may consider
\[
|\eta',\theta\rangle=|\eta\rangle'\otimes|\zeta\rangle=S(\eta',\zeta)|\phi_{0}\rangle\otimes|\psi_{0}\rangle,
\]
where
\[
S(\eta',\theta)= S'(\eta)\otimes S(\theta)=\exp\left[\frac{1}{2}\big(\eta
{a^\dag}^{2}-\ {\eta^*}\tilde{a}^2\big)\right]\otimes\exp\left[\frac{1}{2}\big(\theta
\bar{b}^{2}-\ \bar{\theta}b^2\big)\right].
\]
Also
\[
|\eta,\widetilde{\theta}\rangle=|\eta\rangle\otimes\widetilde{|\zeta\rangle}=S(\eta,\widetilde{\zeta})|\psi_{0}\rangle\otimes|\phi_{0}\rangle,
\]
where
\[
S(\eta,\widetilde{\theta})= S(\eta)\otimes
\widetilde{S(\theta)}=\exp\left[\frac{1}{2}\big(\eta {a^\sharp}^{2}-
{\eta^*}a^2\big)\right]\otimes\exp\left[\frac{1}{2}\big(\theta \bar{c}^{2}-
\bar{\theta}c^2\big)\right].
\]
Finally, using the squeezed state for dual spaces the forthcoming
super squeezed state can be obtained
\[
|\eta',\widetilde{\theta}\rangle=|\eta\rangle'\otimes\widetilde{|\zeta\rangle}=S(\eta',\widetilde{\zeta})|\phi_{0}\rangle\otimes|\phi_{0}\rangle,
\]
here the squeezing operator is
\[
S(\eta',\widetilde{\theta})= S'(\eta)\otimes
\widetilde{S(\theta)}=\exp\left[\frac{1}{2}\big(\eta {a^\dag}^{2}-
{\eta^*}\tilde{a}^2\big)\right]\otimes\exp\left[\frac{1}{2}\big(\theta \bar{c}^{2}-
\bar{\theta}c^2\big)\right].
\]

\section{Entanglement}\label{section5}

In this section, the entanglement of the pseudo-Hermitian coherent
states is investigated. In fact, this is the generalization of~\cite{Naj5,Naj3}, to the states studied in this work. In those
references, considering a  proper linear combination of the tensor
product of the multi-level coherent states as
\[
|\psi\rangle=\sum_{i_1,i_2,\dots,i_n}f_{i_1,i_2,\dots,i_n}|{\theta}_{i_1}\rangle|{\theta}_{i_2}\rangle\cdots |{\theta}_{i_n}\rangle,
\]
and applying an appropriate weight function, it has been shown that,
it is possible to get the right hand side of the following equation
to be an entangled state, like Bell, cluster type, {\bf{GHZ}} and
{\bf{W}} states, i.e.
\[
\int  d{\theta}_{i_1} d{\theta}_{i_2}\cdots
d{\theta}_{i_n}w({\theta}_{i_1},\dots,{\theta}_{i_n})|\psi\rangle
=|\gamma\rangle,
\]
where $ w({\theta}_{i_1},\dots ,{\theta}_{i_n})$  is a proper weight
function, and the state $|\gamma\rangle$, is the entangled state
that we were going to obtain. The weight function is not unique and
of course for a given state there may not be such a function at all.
Pursuing this method, one can see that it is possible to construct
entangled pseudo-Hermitian states. Since for the pseudo-Hermitian
systems we deal with two sets of basis and consequently two set of
coherent states, the possibility of superposing of the states is
more than Hermitian coherent states.

\subsection{Multi-qubit states}\label{section5.1}

Considering (\ref{b,bbar}) and (\ref{c,cbar}) for two level system
($q^2=-1$) we get
\[
b \bar{b}+\bar{b}b=1\qquad \textrm{and}  \qquad
c\bar{c}+\bar{c}c=1.
\]
For a two level system para-Grassmann variables reduce to usual
Grassman variables by the commutation relation
\[
\{{\theta}_{i},{\theta}_{j}\}=0 \quad \Longrightarrow \quad
{\theta}_{i}^2=0.
\]
Thus, the $q$-deformed coherent states based on these operators become
\[
 |\theta\rangle=|\psi_{0}\rangle-\theta|\psi_{1}\rangle, \qquad
  \widetilde{|\theta\rangle}=|\phi_{0}\rangle-\theta|\phi_{1}\rangle.
\]
Now using these states we can construct two level entangled states.
For example, up to the normalization factors, the following
pseudo-Hermitian version of the Bell states can be obtained. Since
these states remind the well-known Bell states, one may call them
pseudo-Hermitian Bell states (PBell states)
\begin{gather*}
\int d\theta|\mp\theta\rangle|-\theta\rangle=|\psi_{0}\rangle
|\psi_{1}\rangle\pm|\psi_{1}\rangle |\psi_{0}\rangle,
\\
\int d\theta
\widetilde{|\mp\theta\rangle}\widetilde{|-\theta\rangle}=
|\phi_{0}\rangle |\phi_{1}\rangle\pm|\phi_{1}\rangle
|\phi_{0}\rangle,
\\
\int d\theta {|\mp\theta\rangle}\widetilde{|-\theta\rangle}=
|\psi_{0}\rangle |\phi_{1}\rangle\pm|\psi_{1}\rangle
|\phi_{0}\rangle,
\\
\int d\theta \widetilde{|\mp\theta\rangle}{|-\theta\rangle}=
|\phi_{0}\rangle |\psi_{1}\rangle\pm|\phi_{1}\rangle
|\psi_{0}\rangle.
\end{gather*}
Also we have
\begin{gather*}
\int
d\theta_2d\theta_1(1+\theta_1\theta_2)|\theta_1\rangle|\pm\theta_2\rangle=|\psi_{0}\rangle
|\psi_{0}\rangle\pm|\psi_{1}\rangle |\psi_{1}\rangle,
\\
\int
d\theta_2d\theta_1(1+\theta_1\theta_2)\widetilde{|\theta_1\rangle}\widetilde{|\mp\theta_2\rangle}=|\phi_{0}\rangle
|\phi_{0}\rangle\pm|\phi_{1}\rangle |\phi_{1}\rangle,
\\
\int
d\theta_2d\theta_1(1+\theta_1\theta_2){|\theta_1\rangle}\widetilde{|\mp\theta_2\rangle}=|\psi_{0}\rangle
|\phi_{0}\rangle\pm|\psi_{1}\rangle |\phi_{1}\rangle,
\\
\int
d\theta_2d\theta_1(1+\theta_1\theta_2)\widetilde{|\theta_1\rangle}{|\mp\theta_2\rangle}=|\phi_{0}\rangle
|\psi_{1}\rangle\pm|\phi_{1}\rangle |\psi_{0}\rangle.
\end{gather*}
It is notable that the normalization factor can be included in
weight function. One gets the general form of the {\bf{W}} state as
follows
\begin{gather*}
\int
d\theta(\frac{-1}{\sqrt{N}})\underbrace{|\theta\rangle|\theta\rangle \cdots |\theta\rangle}_{n\
{\rm times}}=|{\bf{W}}^{(n)}_1\rangle,
\\
|{\bf{W}}^{(n)}_1\rangle=\frac{1}{\sqrt{N}}(|\psi_{1}\psi_{0}\psi_{0}\cdots \psi_{0}\rangle+|\psi_{0}\psi_{1}\psi_{0}\cdots \psi_{0}\rangle+
\cdots +|\psi_{0}\cdots \psi_{0}\psi_{0}\psi_{1}\rangle),
\end{gather*}
here, $\frac{1}{\sqrt{N}}$ is the normalization factor of the state
$|\psi_{1}\psi_{0}\psi_{0}\cdots \psi_{0}\rangle+|\psi_{0}\psi_{1}\psi_{0}\cdots \psi_{0}\rangle+
\cdots +|\psi_{0}\cdots \psi_{0}\psi_{0}\psi_{1}\rangle$. We can obtain the
other states of this family like above. For instance,
\begin{gather*}
\int
d\theta(\frac{-1}{\sqrt{N}})\underbrace{\widetilde{|\theta\rangle}|\theta\rangle\cdots |\theta\rangle}_{n\
{\rm times}}=|{\bf{W}}^{(n)}_2\rangle,
\\
|{\bf{W}}^{(n)}_2\rangle=\frac{1}{\sqrt{N}}(|\phi_{1}\psi_{0}\psi_{0}\cdots \psi_{0}\rangle+|\phi_{0}\psi_{1}\psi_{0}\cdots \psi_{0}\rangle+
\cdots +|\phi_{0}\cdots \psi_{0}\psi_{0}\psi_{1}\rangle).
\end{gather*}
Likewise, we can obtain these states
\begin{gather*}
|{\bf{W}}^{(n)}_3\rangle=\frac{1}{\sqrt{N}}(|\phi_{1}\phi_{0}\psi_{0}\cdots \psi_{0}\rangle+|\phi_{0}\phi_{1}\psi_{0}\cdots \psi_{0}\rangle+
\cdots +|\phi_{0}\cdots \psi_{0}\psi_{0}\psi_{1}\rangle),
\\
|{\bf{W}}^{(n)}_4\rangle=\frac{1}{\sqrt{N}}(|\phi_{1}\phi_{0}\phi_{0}\cdots \psi_{0}\rangle+|\phi_{0}\phi_{1}\phi_{0}\cdots \psi_{0}\rangle+
\cdots +|\phi_{0}\cdots \psi_{0}\psi_{0}\psi_{1}\rangle),
\\
|{\bf{W}}^{(n)}_k\rangle=\frac{1}{\sqrt{N}}(|\phi_{1}\phi_{0}\phi_{0}\cdots \phi_{0}\rangle+|\phi_{0}\phi_{1}\phi_{0}\cdots \phi_{0}\rangle+
\cdots +|\phi_{0}\cdots \phi_{0}\phi_{0}\phi_{1}\rangle).
\end{gather*}
Likewise, we can construct the general forms of the {\bf{GHZ}} state
as follows
\begin{gather*}
\int
d\theta_{1}d\theta_{2}\cdots d\theta_{n}w|\theta_{n}\rangle|\theta_{n-1}\rangle\cdots |\theta_{1}\rangle=\frac{1}{\sqrt{2}}
(|\psi_{0}\psi_{0}\cdots \psi_{0}\rangle+|\psi_{1}\psi_{1}\cdots \psi_{1}\rangle)=|{\bf{GHZ}}^{(n)}_1\rangle,
\end{gather*}
and
\begin{gather*}
\int
d\theta_{1}d\theta_{2}\cdots d\theta_{n}w\widetilde{|\theta_{n}\rangle}|\theta_{n-1}\rangle\cdots |\theta_{1}\rangle=\frac{1}{\sqrt{2}}
(|\phi_{0}\psi_{0}\cdots \psi_{0}\rangle+|\phi_{1}\psi_{1}\cdots \psi_{1}\rangle)=|{\bf{GHZ}}^{(n)}_2\rangle,
\end{gather*}
and also
\begin{gather*}
\int
d\theta_{1}d\theta_{2}\cdots d\theta_{n}w\widetilde{|\theta_{n}\rangle}\widetilde{|\theta_{n-1}\rangle}\cdots \widetilde{|\theta_{1}\rangle}=\frac{1}{\sqrt{2}}
(|\phi_{0}\phi_{0}\cdots \phi_{0}\rangle+|\phi_{1}\phi_{1}\cdots \phi_{1}\rangle)=|{\bf{GHZ}}^{(n)}_k\rangle,
\end{gather*}
 where the weight function $w$ for the above states is
\[
w=\frac{1}{\sqrt{2}}\big((-1)^{[\frac{n}{2}]}+\theta_{n}\theta_{n-1}\cdots \theta_{1}\big).
\]
 It is notable that, when the
pseudo-Hermitian system reduces to the hermitian one ($\eta=1$), the
pseudo-Hermitian entangled states reduces to the well-known
maximally entangled states of the Hermitian systems, like maximally
entangled  Bell, {\bf{W}} and  {\bf{GHZ}}. So fore instance  the
states~$|{\bf{W}}^{(n)}_i\rangle$ reduce to just one maximally
entangled state.

\subsection{Multi-qutrit states}\label{section5.2}

Using three level coherent states, it would be possible to construct
entangled qutrit states. Since for three level systems the operators
$b^3$ and $\bar{b}^3$  vanish, the expansion of the equation
(\ref{coh-gen}) reduces to the following three level coherent state
\[
|\theta\rangle=|\psi_{0}\rangle+\frac{q^2}{\sqrt{[1]}!}\theta|\psi_{1}\rangle+\frac{\theta^2}{\sqrt{[2]}!}|\psi_{2}\rangle.
\]
Obviously, the para-Grassmann number $\theta$ is nilpotent, where
$\theta^3=0$, and also $(q^2)^3=1$. Now this coherent state will be
used to construct some  entangled qutrit states. Here we introduce
the construction of generalized Bell states for pseudo-Hermitian
systems. The generalized Bell states for three level Hermitian
systems are
\begin{gather}\label{gener-Bell1}
|\psi_\pm\rangle=\frac{1}{\sqrt{3}}(|00\rangle\pm|11\rangle+|22\rangle),
\\
\label{gener-Bell2}
|\varphi_\pm\rangle=\frac{1}{\sqrt{3}}(|02\rangle\pm|11\rangle+|20\rangle).
\end{gather}
Here, we try to create the pseudo-Hermitian version of these states
by using three level para-Grassmannian  coherent states
\begin{gather*}
\int d\bar{\theta}d\theta
w|\theta\rangle|\bar{\theta}\rangle=|\psi_{0}\rangle
|\psi_{0}\rangle\pm|\psi_{1}\rangle
|\psi_{1}\rangle+|\psi_{2}\rangle |\psi_{2}\rangle,
\\
\int d\bar{\theta}d\theta
w\widetilde{|\theta\rangle}|\bar{\theta}\rangle=|\phi_{0}\rangle
|\psi_{0}\rangle\pm|\phi_{1}\rangle
|\psi_{1}\rangle+|\phi_{2}\rangle |\psi_{2}\rangle,
\\
\int d\bar{\theta}d\theta
w|\theta\rangle\widetilde{|\bar{\theta}\rangle}=|\psi_{0}\rangle
|\phi_{0}\rangle \pm|\phi_{1}\rangle
|\psi_{1}\rangle+|\phi_{2}\rangle |\psi_{2}\rangle,
\\
\int d\bar{\theta}d\theta
w\widetilde{|\theta\rangle}\widetilde{|\bar{\theta}\rangle}=|\phi_{0}\rangle
|\phi_{0}\rangle \pm|\phi_{1}\rangle
|\phi_{1}\rangle+|\phi_{2}\rangle |\phi_{2}\rangle,
\end{gather*}
where, for all above states
\[
w=\frac{\theta^2\bar{\theta}^2}{{[2]}!}\pm\frac{\theta\bar{\theta}}{{[1]}!}+q^2.
\]
One may compare the above pseudo-Hermitian entangled states, with
 maximally entangled generalized Bell state~(\ref{gener-Bell1}).
Note that in the case $\eta=1$ $(\phi_{i}=\psi_{i})$, all the states
reduce to~(\ref{gener-Bell1}). Moreover, we have the following
pseudo-Hermitian entangled states
\begin{gather*}
\int d\bar{\theta}d\theta
w|\theta\rangle|\bar{\theta}\rangle=|\psi_{0}\rangle
|\psi_{2}\rangle\pm|\psi_{1}\rangle
|\psi_{1}\rangle+|\psi_{2}\rangle |\psi_{0}\rangle,
\\
\int d\bar{\theta}d\theta
w\widetilde{|\theta\rangle}|\bar{\theta}\rangle=|\phi_{0}\rangle
|\psi_{2}\rangle\pm|\phi_{1}\rangle
|\psi_{1}\rangle+|\phi_{2}\rangle |\psi_{0}\rangle,
\\
\int d\bar{\theta}d\theta
w|\theta\rangle\widetilde{|\bar{\theta}\rangle}=|\psi_{0}\rangle
|\phi_{2}\rangle \pm|\phi_{1}\rangle
|\psi_{1}\rangle+|\phi_{2}\rangle |\psi_{0}\rangle,
\\
\int d\bar{\theta}d\theta
w\widetilde{|\theta\rangle}\widetilde{|\bar{\theta}\rangle}=|\phi_{0}\rangle
|\phi_{2}\rangle \pm|\phi_{1}\rangle
|\phi_{1}\rangle+|\phi_{2}\rangle |\phi_{0}\rangle,
\end{gather*}
where
\[
w=\frac{\theta^2}{\sqrt{[2]!}}\pm\frac{\theta\bar{\theta}}{{[1]}!}+q^2\frac{\bar{\theta}^2}{\sqrt{[2]!}}.
\]
These states can be compared with maximally entangled generalized
Bell state (\ref{gener-Bell2}). Of course the weight function may be
chosen in a way that we get entangled states in the subspaces. For
instance,
\begin{gather*}
\int d\bar{\theta}d\theta
\left(\frac{\theta^2\bar{\theta}^2}{{[2]}!}+q^2\right)|\theta\rangle|\bar{\theta}\rangle=|\psi_{0}\rangle
|\psi_{0}\rangle+|\psi_{2}\rangle |\psi_{2}\rangle,
\\
\int d\bar{\theta}d\theta
\left(\frac{\theta^2\bar{\theta}^2}{{[2]}!}+q^2\right)\widetilde{|\theta\rangle}\widetilde{|\bar{\theta}\rangle}=|\phi_{0}\rangle
|\phi_{0}\rangle +|\phi_{2}\rangle |\phi_{2}\rangle.
\end{gather*}

\subsection{Multi-qudit entangled states}\label{section5.3}

Now, using the coherent state obtained in (\ref{coh-gen}), the
entangled states for $n$ level pseudo-Hermitian systems will be
considered. To this aim starting with the coherent state
\[
|\theta\rangle=\sum_{n=0}^{p}
\frac{{q}^{^{{n(n+1)}}}}{\sqrt{[n]!}}  \theta ^{n}  |\psi_n\rangle,
\]
the product of two coherent states with para-Grassmann numbers
$\theta$ and $\bar{\theta}$ reads
\[
|\theta\rangle|\bar{\theta}\rangle=\sum_{i,j=0}^{p}c_{ij}
\theta^i\bar{\theta}^j|\psi_i\rangle|\psi_j\rangle,
\]
where
\[
c_{ij}=\frac{q{^{{(j-i)^2+(i+j)}}}}{\sqrt{[i]![j]!}}.
\]
For instance,  we may determine the weight function in a way that
after integration the obtained state becomes
\begin{gather}\label{gen2}
\int d\bar{\theta}d\theta
w|\theta\rangle|\bar{\theta}\rangle=\frac{1}{\sqrt{N}}\sum_{i=0}^{p}
|\psi_i\rangle|\psi_i\rangle.
\end{gather}
Replacing the explicit form of the coherent states we get
\begin{gather}\label{gen1}
\int d\theta_1d\theta_2w\sum_{i,j=0}^{n-1}c_{ij}
\theta_1^i\theta_2^j|\psi_i\rangle|\psi_j\rangle
=\frac{1}{\sqrt{N}}\sum_{i=0}^{p} |\psi_i\rangle|\psi_i\rangle.
\end{gather}
Taking the general form of the weight function
\[
w=\sum_{k,l=0}^{p}w_{k,l}\bar{\theta}^{k}\theta^{l},
\]
and putting this weight in (\ref{gen1}) and taking into account the
quantization  and the integration rules of para-Grassmannian
variables  we get
\begin{gather}
\sum_{k,l=0}^{p}\sum_{i,j=0}^{p}c_{ij}w_{k,l}\int
d\bar{\theta}d\theta\bar{\theta}^{k}\theta^{l+i}
\bar{\theta}^j|\psi_i\rangle|\psi_j\rangle
=\frac{1}{\sqrt{N}}\sum_{i=0}^{p} |\psi_i\rangle|\psi_i\rangle,
\nonumber\\
 \sum_{k,l=0}^{p}\sum_{i,j=0}^{p}c_{ij}w_{k,l}\bar{q}^{2k(l+i)}\int
d\bar{\theta}d\theta\theta^{l+i}\bar{\theta}^{k+j}
|\psi_i\rangle|\psi_j\rangle =\frac{1}{\sqrt{N}}\sum_{i=0}^{p}
|\psi_i\rangle|\psi_i\rangle,
\nonumber\\
\label{qudit1}
 \sum_{k,l=0}^{p}\sum_{i,j=0}^{p}c_{ij}w_{k,l}\bar{q}^{2k(l+i)}
\delta^{l+i}_{p}\delta^{k+j}_{p}[p]!|\psi_i\rangle|\psi_j\rangle
=\frac{1}{\sqrt{N}}\sum_{i=0}^{p} |\psi_i\rangle|\psi_i\rangle,
\end{gather}
where the symbol  $\delta^i_{j}$ is the usual Kronecker delta. Note
that
\begin{gather}\label{qudit2}
\delta^{l+i}_{p}\delta^{k+j}_{p}\neq0 \quad \Longrightarrow \quad
{l+i}=p=k+j.
\end{gather}
From the right hand side if the the equation (\ref{qudit1}) it is
clear that the terms with $i\neq j$ have to vanish in the left hand
side of the equation, which due to (\ref{qudit2}) we get $w_{k,l}=0$
for $k\neq l$.
 With this description (\ref{qudit1}) gives
\[
\sum_{i=0}^{p}c_{ii}w_{p-i,p-i}\bar{q}^{^{2p(p-i)}}[p]!
|\psi_i\rangle|\psi_i\rangle =\frac{1}{\sqrt{N}}\sum_{i=0}^{p}
|\psi_i\rangle|\psi_i\rangle.
\]
Thus,
\[
w=\frac{1}{\sqrt{N}}\sum_{k=0}^{p}c_{(p-k),(p-k)}^{-1}\frac{{q}^{2pk}}{[p]!}\bar{\theta}^{k}\theta^{k}.
\]
With this weight (\ref{gen2}) holds. Using this weight function the
below entangled states can be constructed
\[
\int d\bar{\theta}d\theta
w\widetilde{|\theta\rangle}\widetilde{|\bar{\theta}\rangle}=\frac{1}{\sqrt{N}}\sum_{i=0}^{p}
|\phi_i\rangle|\phi_i\rangle.
\]
Of course the normalization factors $\big(\frac{1}{\sqrt{N}}\big)$ of the
above states may  not be the same.

\section{Conclusion}\label{section6}

In conclusion, the $q$-deformed oscillator was generalized to
pseudo-Hermitian systems and some of its important properties was
studied. Introducing annihilation and creation operators for this
system the new pseudo-Hermitian coherent and squeezed states were
investigated. The over-completeness property of the PGPHCSs
examined. Also the stability of  coherent and squeezed states were
discussed. The pseudo-Hermitian supercoherent states as the product
of a pseudo-Hermitian bosonic coherent state and a para-Grassmannian
pseudo-Hermitian coherent state was introduced. This def\/inition also
was developed to def\/ine pseudo-Hermitian supersqueezed states. It
was shown that, for $q$-oscillator algebra of~$k+1$ degree of nilpotency
based on the changed ladder operators, def\/ined in here, we can
obtain deformed~$SU_{q^2}(2)$ and~$SU_{q^{2k}}(2)$ and also
$SU_{q^{2k}}(1,1)$. Finally, the entanglement of multi-level
para-Grassmannian pseudo-Hermitian coherent state was considered.
This was done by choosing an appropriate weight function, and
integrating over tensor product of PGPHCSs. Thus, a relation between
PGPHCSs and quantum entanglement was established. The entangled
pseudo-Hermitian qubit states based on two level coherent states was
given, and it was generalized for qutrit and qudit cases in general.

\pdfbookmark[1]{References}{ref}
\LastPageEnding

\end{document}